\documentclass[12pt,a4paper]{article}

\usepackage{amsmath,amssymb,graphicx,subcaption}

\def\be{\begin{equation}}
\def\ee{\end{equation}}
\def\d{\text{d}}
\def\Spo{\Sigma_{+0}}
\def\Sno{\Sigma_{-0}}
\def\Sp{\Sigma_{+}}
\def\Sm{\Sigma_{-}}

\begin{document}

\begin{center}
{\Large{\bf Cylindrical spikes}}

\

{\bf M Z A Moughal, W C Lim}

\

Department of Mathematics, University of Waikato, Private Bag 3105, Hamilton 3240, New Zealand

\

moughalzubair@gmail.com, wclim@waikato.ac.nz
\end{center}

\begin{abstract} 
The Geroch/Stephani transformation is a solution-generating transformation, and may generate spiky solutions. The spikes in solutions generated so far are either early-time 
permanent spikes or transient spikes. We want to generate a solution with a late-time permanent spike. We achieve this by applying Stephani's transformation with the rotational Killing vector 
field of the locally rotationally symmetric Jacobs solution. The late-time permanent spike occurs along the cylindrical axis. 
Using a mixed Killing vector field, the generated solution also features a rich variety of transient structures. 
We introduce a new technique to analyse these structures. Our findings lead us to discover a transient behaviour, which we call the overshoot transition.
These discoveries compel us to revise the description of transient spikes.
\end{abstract} 


\section{Introduction}

Spikes are small-scale spatial structures that form and then either remain there (permanent spikes) or disappear (transient spikes).
Spikes were discovered incidentally by Berger and Moncrief~\cite{art:BergerMoncrief1993}, whose original goal was to understand the nature of generic singularities.
The well-known BKL conjecture by
Lifshitz, Khalatnikov and Belinskii~\cite{art:LK63,art:BKL1970,art:BKL1982} 
are heuristic arguments that the approach to generic spacelike singularities is vacuum dominated, local and oscillatory.
The non-local nature of spikes brings the the local nature of the conjecture into doubt.
See~\cite{art:HeinzleUgglaLim2012} for a comprehensive introduction.
Since 2012 the focus has shifted to the possible role of spikes in the formation of large-scale structures as the Universe expands~\cite{art:ColeyLim2012,art:LimColey2014,art:ColeyLim2014}.

In a series of four papers~\cite{art:Lim2015,art:ColeyLim2016,art:Coleyetal2016,art:Gregorisetal2017}, 
spiky solutions were generated using Geroch's and Stephani's transformations \cite{art:Geroch1971,art:Geroch1972,art:Stephani1988} in order to study the dynamics of spikes. 
One of the goals is to undestand the role of spikes in the formation of filamentary structures.
The spikes in solutions generated so far are either early-time permanent spikes or transient spikes.
In the conclusion section of~\cite{art:Gregorisetal2017}, the authors hoped to generate solutions with a late-time permanent spike, which is more suitable for formation of permanent structures.
This became the initial goal for the PhD thesis of the first author~\cite{thesis:Moughal2021}.
The goal was achieved by applying Stephani's transformation with the rotational Killing vector field (KVF) of the locally rotationally symmetric (LRS) Jacobs solution.
With a mixed KVF, the generated solution unexpectedly features a rich variety of transient structures. 
A new technique was introduced to analyse these structures, leading to the discovery of a transient behaviour, which we call the overshoot transition,
and also leading to the re-examination of the definition of transient spikes.
This paper is an abridged version of the thesis.

\section{The metric and the Iwasawa frame}

Assume zero vorticity (zero shift). The spatial metric components are given by the formula $g_{ij} = e^\alpha{}_i e^\beta{}_j \delta_{\alpha\beta}$,
where Roman indices $i$, $j=1..3$ are spatial coordinate indices, and Greek indices $\alpha$, $\beta = 1..3$ are spatial orthonormal frame indices.
The Iwasawa frame~\cite{art:HeinzleUgglaRohr2009} is a choice of orthonormal frame that makes $e^\alpha{}_i$ (and equivalently $e_\alpha{}^i$) upper triangular, as follows.
The frame coefficients $e^\alpha{}_i$ simplify from 9 components to 6 components, represented by $b^1$, $b^2$, $b^3$, $n_1$, $n_2$ and $n_3$.

\begin{align}
e^\alpha{}_i = \left( \begin{array}{ccc}
e^1{}_1 & e^1{}_2 & e^1{}_3 \\	
e^2{}_1	& e^2{}_2 & e^2{}_3 \\
e^3{}_1	& e^3{}_2 & e^3{}_3
\end{array} \right)
&= \left( \begin{array}{ccc}
e^{-b^1} & 0 & 0 \\
0 & e^{-b^2} & 0 \\
0 & 0 & e^{-b^3}
\end{array} \right)
\left( \begin{array}{ccc}
1 & n_1 & n_2 \\
0 & 1   & n_3 \\
0 & 0   & 1
\end{array} \right)
\notag\\
&= \left( \begin{array}{ccc}
e^{-b^1} & e^{-b^1} n_1 & e^{-b^1} n_2 \\
0 & e^{-b^2} & e^{-b^2} n_3 \\
0 & 0 & e^{-b^3}
\end{array} \right)
\end{align}
\begin{align}
e_\alpha{}^i = \left( \begin{array}{ccc}
e_1{}^1 & e_2{}^1 & e_3{}^1 \\
e_1{}^2 & e_2{}^2 & e_3{}^2 \\
e_1{}^3 & e_2{}^3 & e_3{}^3   
\end{array} \right)
&= \left( \begin{array}{ccc} 
1 & -n_1	& n_1 n_3 - n_2 \\
0 & 1   & -n_3 \\
0 & 0  	& 1
\end{array} \right)
\left( \begin{array}{ccc}
e^{b^1} & 0 & 0 \\
0 & e^{b^2} &	0 \\
0 & 0 &	e^{b^3}
\end{array} \right)
\notag\\
&= \left( \begin{array}{ccc}
e^{b^1} & -e^{b^2} n_1 & e^{b^3} (n_1 n_3 - n_2) \\
0 & e^{b^2} & -e^{b^3} n_3 \\
0 & 0 & e^{b^3}
\end{array} \right)
\end{align}
The frame derivative operators $\mathbf{e}_0 = N^{-1} \partial_0$, $\mathbf{e}_\alpha = e_\alpha{}^i \partial_i$ in the Iwasawa frame are
\begin{align}
\mathbf{e}_0 &= \frac{1}{N} \partial_0
\\
\mathbf{e}_1 &= e^{b^1} \partial_1
\\
\mathbf{e}_2 &= e^{b^2} [ -n_1 \partial_1 + \partial_2 ]
\\
\mathbf{e}_3 &= e^{b^3} [ (n_1 n_3 - n_2) \partial_1 - n_3 \partial_2 + \partial_3 ].
\end{align}

In the Iwasawa frame, the metric components in terms of the $b$'s and $n$'s are given by
\begin{align}
g_{00} &= -N^2
\\
g_{11} &= e^{-2b^1},\quad g_{12} = e^{-2b^1} n_1,\quad g_{13} = e^{-2b^1} n_2
\\
g_{22} &= e^{-2b^2} + e^{-2b^1} n_1^2,\quad g_{23} =  e^{-2b^1} n_1 n_2 + e^{-2b^2} n_3
\\
g_{33} &= e^{-2b^3} + e^{-2b^1} n_2^2 + e^{-2b^2} n_3^2.
\end{align}

If the metric is given, we can compute the $b$'s and $n$'s as follows.
\begin{align}
b^1 &= -\tfrac12 \ln g_{11}
\\
n_1 &= \frac{g_{12}}{g_{11}}
\\
n_2 &= \frac{g_{13}}{g_{11}}
\\
b^2 &= -\tfrac12 \ln ( g_{22} - g_{12} n_1 )
\\
n_3 &= (g_{23}-g_{12} n_2) e^{2b^2}
\\
b^3 &= -\tfrac12 \ln ( g_{33} - g_{13} n_2 - e^{-2b^2} n_3^2 ).
\end{align}

\section{Geroch's and Stephani's transformations}

Consider a solution $g_{ab}$ of the vacuum Einstein's field equations with a KVF $\xi^a$. 
Geroch's transformation~\cite{art:Geroch1971,art:Geroch1972} (see also \cite[Section 10.3]{book:Exactsol2002}) is an algorithm for generating new solutions, by exploiting the KVF $\xi^a$.
The algorithm involves solving the following partial differential equations
\begin{gather}
\label{omega_PDE}
\nabla_a \omega =\varepsilon_{abcd}\xi ^b\nabla^c \xi^d,
\\
\label{alpha_PDE}
\nabla_{[a}\alpha_{b]} =\frac{1}{2}\varepsilon_{abcd} \nabla^c \xi^d,\quad
\xi^a \alpha_a =\omega,
\\
\label{beta_PDE}
\nabla_{[a}\beta_{b]}=2\lambda \nabla_a \xi_b + \omega \varepsilon_{abcd}  \nabla^c \xi^d,
\quad
\xi^a \beta_a =\lambda^2 +\omega^2 -1
\end{gather}
for $\omega$, $\alpha_a$ and $\beta_a$, where $\lambda=\xi^a \xi_a$.
Next, define $\tilde{\lambda}$ and $\eta_a$ as
\begin{align}
\tilde{\lambda}&=\lambda \Big[(\cos\theta-\omega\sin\theta)^2 +\lambda^2 \sin^2\theta\Big]^{-1},
\\
\label{eta}
\eta_a &=\tilde{\lambda}^{-1} \xi_a +2 \alpha_a \cos\theta\sin\theta-\beta_a \sin^2\theta,
\end{align} 
for any constant $\theta$. 
Then the new metric is given by
\be
\tilde{g}_{a b}=\frac{\lambda}{\tilde{\lambda}}(g_{a b}-\lambda^{-1} \xi_a\xi_b)+\tilde{\lambda} \eta_a \eta_b.
\ee
This new metric is again a solution of the vacuum Einstein's field equations with the same KVF.
$\theta=0$ gives the trivial transformation $\bar{g}_{a b} = g_{ab}$.
 
Notice from (\ref{eta}) that $\alpha_a$ appears in the new metric only through $\eta_a$, and if $\theta$ is chosen to be $\pi/2$ then $\alpha_a$ does not appear at all. 
We shall exploit this simplification. In this case the new metric simplifies to
\be
\tilde{g}_{a b}= F g_{a b} + \frac{\lambda}{F}\beta_a \beta_b - \xi_a \beta_b - \beta_a \xi_b,
\ee
where
\be
	F = \lambda^2+\omega^2.
\ee

Stephani~\cite{art:Stephani1988} generalised Geroch's transformation to the case of comoving stiff fluid
if the KVF is spacelike (and to the case of perfect fluid with equation of state $p=-\rho/3$ if the KVF is timelike, which we do not study here). 
The algorithm is the same as before, with the new stiff fluid density given by
\be
\tilde{\rho} = \frac{\rho}{F}.
\ee

Before applying Geroch's or Stephani's transformation, we set up the coordinates such that the KVF to be used has the form
\begin{equation}
	\xi^a = (0,\ 1,\ 0,\ 0),
\end{equation}
to adapt to the Iwasawa frame for simplicity.

In simpler cases, if the seed metric has the form
\begin{equation}
\label{gmetric1}
        g_{ab} = \begin{bmatrix}
                -N^2    &       0       &       0       &       0       \\
                0       &       g_{11}  &       g_{12}  &       0  \\
                0       &       g_{12}  &       g_{22}  &       0  \\
                0       &       0	&       0	&       g_{33}  
                \end{bmatrix},
\end{equation}
i.e. if $n_2=0=n_3$,
then the generated metric has the form
\begin{equation}
\label{gmetric2}
        \tilde{g}_{ab} = \begin{bmatrix}
                -FN^2    &       0       	&       0       &       0       \\
                0       &       \tilde{\lambda}  &       g_{12}-\beta_2 \tilde{\lambda}  &       0  \\
                0       &       \tilde{g}_{12}  &       Fg_{22}-2g_{12}\beta_2+\beta_2^2\tilde{\lambda}  &       0  \\
                0       &        0		&       0  &       Fg_{33} 
                \end{bmatrix}.
\end{equation}
Expressing the metric $\tilde{g}_{ab}$ in (\ref{gmetric2}) in $b$'s and $n$'s gives
\begin{align}
\label{Geroch_bn_kzero_first}
\tilde{N}&=N\sqrt{F}\\
\tilde{b}^1&=b^1+\frac12\ln F \\
\tilde{b}^2&=b^2-\frac12\ln F \\
\tilde{b}^3&=b^3-\frac12\ln F \\
\tilde{n}_{1}&=n_{1}F-\beta_{2}\\
\tilde{n}_{2}&=0\\
\label{Geroch_bn_kzero_last}
\tilde{n}_{3}&=0.
\end{align}

\section{LRS Jacobs seed solution with rotational KVF}

Spacelike KVFs can be classified into two kinds -- translational and rotational.
The four papers~\cite{art:Lim2015,art:ColeyLim2016,art:Coleyetal2016,art:Gregorisetal2017} used a linear combination of translational KVFs. 
In this paper, we will use a linear combination of KVFs that includes a rotational KVF. 
Stephani's transformation requires the matter to be a stiff fluid, so we start by looking at locally rotationally symmetric (LRS) solutions with a stiff fluid. 
The simplest such solution is the flat FLRW solution, but it does not generate as much structure as the next simplest solution, the LRS Jacobs (Bianchi type I) solution, 
which we shall use as the seed solution.

The Jacobs solution~\cite{art:Jacobs1968}\cite[page 189]{book:WainwrightEllis1997} is given by the line element 
\be
	\d s^2=-\d t^2+t^{2p_{1}}\d x^2+t^{2p_{2}}\d y^2+t^{2p_{3}}\d z^2,
\ee
where the coordinates are $(t,x,y,z)$,
and
\begin{align}
p_1 &= \frac13(1+\Spo + \sqrt3\Sno),\\
p_2 &= \frac13(1+\Spo - \sqrt3\Sno),\\
p_3 &= \frac13(1-2\Spo).
\end{align}
The non-zero Hubble-normalised shear components~\cite[Sections 1.1.3, 6.1.1]{book:WainwrightEllis1997} are $\Spo$ and $\Sno$, and they are constant, with $\Spo^2+\Sno^2\leq1$.
The comoving stiff fluid has pressure $p$ and density $\rho$ given by
\be
p = \rho = \frac{1-\Spo^2-\Sno^2}{3t^2}.
\ee

To impose the LRS condition, it is simplest to set $\Sno=0$, so the parameter $\Spo$ takes values from $-1$ to $1$.
$\Spo=-1$ gives the LRS Kasner solution \cite[page 132]{book:WainwrightEllis1997} with $(p_1,p_2,p_3)=(0,0,1)$ (also known as the Taub form of flat spacetime);
$\Spo=1$ gives the LRS Kasner solution with $(p_1,p_2,p_3)=(\frac23,\frac23,-\frac13)$;
$\Spo=0$ gives the flat FLRW solution with stiff fluid.

The LRS Jacobs solution admits four KVFs, namely
\be
\partial_x,\quad \partial_y,\quad \partial_z,\quad -y \partial_x + x \partial_y,
\ee
where the fourth one is rotational.
We intend to apply Stephani's transformation with the general linear combination of the KVFs:
\be
c_1 \partial_x + c_2 \partial_y + c_3 \partial_z + c_4(-y\partial_x + x\partial_y)
= (c_1-c_4 y) \partial_x + (c_2 + c_4 x)\partial_y + c_3 \partial_z
\ee
Observe that $c_1$ and $c_2$ can be eliminated without loss of generality by a translation in $x$ and $y$ directions. We set $c_4=1$ and $c_3=k$, so the KVF reads
\be
	-y \partial_x + x \partial_y + k \partial_z.
\ee
This KVF forms an Abelian orthogonally transitive (OT) $G_2$ group with exactly one other KVF (namely a linear combination of $\partial_z$ and $-y \partial_x + x \partial_y$). 
By Geroch's theorem~\cite[Appendix B]{art:Geroch1972}, the generated metric will admit an Abelian OT $G_2$ group.

There is a rotational symmetry about the $z$-axis, so we adopt cylindrical coordinates $(r,\psi,z)$, but we want to arrange the coordinates in the following order: $(t,\psi,z,r)$, 
due to the way we adapt the orthonormal frame to the coordinates.
In these coordinates, the KVF reads
\be
	\partial_\psi + k \partial_z.
\ee
We want to simplify the KVF to just $\partial_\psi$ for the application of Stephani's transformation, so we make a further change of coordinates,
by introducing
\be
	Z = z - k\psi.
\ee
Then, in the coordinates $(t,\psi,Z,r)$, the KVF is simply $\partial_\psi$, but
the line element now reads
\begin{align}
\label{seed1}
	\d s^2 &= -\d t^2+(k^2t^{2p_3}+r^2t^{2p_1})\d\psi^2+2kt^{2p_3}\d\psi \d Z +t^{2p_3}\d Z^2+t^{2p_1}\d r^2.
\end{align}
This shall be the seed solution to which we apply Stephani's transformation. It has the simple form (\ref{gmetric1}).

\begin{figure}[t]
	\begin{center}
		\includegraphics[width=8cm]{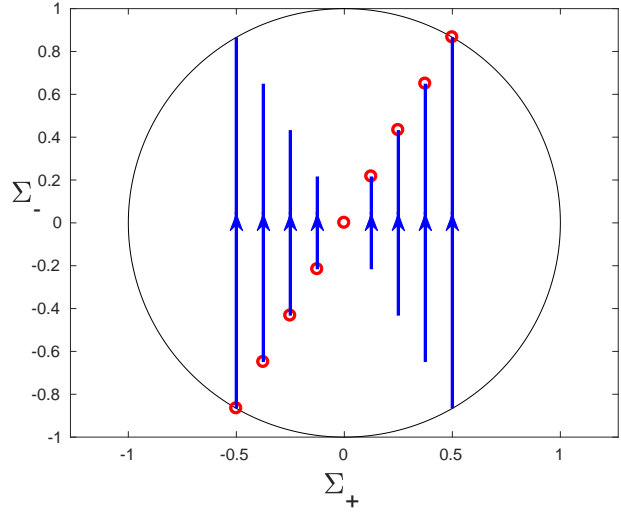}	
	\end{center}
	\caption{State space orbits of the rotated LRS Jacobs solution projected on the $(\Sp,\Sm)$ plane for various values of $\Spo$, assuming $k\neq0$. The $r=0$ orbits are fixed points.
		As $t$ increases, $r\neq0$ orbits move away from these fixed points for $\Spo >0$, and towards these fixed points for $\Spo <0$.}
	\label{seedorbits}
\end{figure}

The state space orbits of a solution, projected onto the $(\Sp,\Sm)$ plane, can provide some insight into the dynamics of the solution.
Recall that $(\Sp,\Sm)$ are defined in terms of the diagonal components of the Hubble-normalised expansion shear as
\be
        \Sp =-\frac{1}{2}\Sigma_{33},\quad \Sm =\frac{\Sigma_{11}-\Sigma_{22}}{2\sqrt3}.
\ee
which gives
\be
\Sp=-\frac12 \Spo,\quad \Sm=\frac{\sqrt3}{2}\left(l-\frac{2-\Spo}{3}\right),\quad l = t(\ln\lambda)_t.
\ee
The solution is undefined at $r=0$ if $k=0$ (coordinate singularity).
It is straightforward to analyse $l$.
If $k=0$ then $l=2p_1$.
If $k\neq0$, then $l=2p_3$ at $r=0$.
For $r\neq0$ write
\be
\label{3.50}
l=p_3(1-\tau)+p_1(1+\tau), \quad\tau=\tanh(\Spo(\ln t)+\ln|k|-\ln r).
\ee
For $\Spo>0$, $l$ goes from $2p_3$ to $2p_1$ as $t$ goes from $0$ (early times) to $\infty$ (late times).
For $\Spo<0$, $l$ goes from $2 p_1$ to $2p_3$. For $\Spo=0$, $l=\frac23$. So $l$ has a simple sigmoid transitional dynamics.
It has a discontinuous limit along $r=0$ (at late times for $\Spo>0$, at early times for $\Spo<0$).
This creates a permanent false spike along the cylindrical axis $r=0$ at late times for $\Spo >0$, and at early times for $\Spo <0$.
The false spike is entirely a coordinate effect, due to the rotating Iwasawa frame.
Figure \ref{seedorbits} shows that state space orbits projected on the $(\Sp,\Sm)$ plane for various values of $\Spo$, assuming $k\neq0$. The $r=0$ orbits are fixed points.
As $t$ increases, $r\neq0$ orbits move away from these fixed points for $\Spo >0$, and towards these fixed points for $\Spo <0$.

\section{The generated cylindrical solution}

We now carry out Stephani's transformation with the general KVF $\partial_\psi$.
We obtain
\be
\lambda = k^2t^{2p_3}+r^2t^{2p_1},\quad
\omega= \dfrac{2k}{1+p_3} t^{1+p_3}+k\Spo r^2+\omega_{0},
\ee
\be
\beta_2 = 2p_1\omega_0r^2+ p_1\Spo kr^4 + \left(\frac{4\omega+2k(1-p_3)r^2}{1+p_3} \right)t^{1+p_3}+k^3 t^{4p_3} + \frac{4k}{(1+p_3)^2} t^{2+2p_3}.
\ee
The generated metric is then given through $b$'s and $n$'s by the formulas (\ref{Geroch_bn_kzero_first})--(\ref{Geroch_bn_kzero_last}), dropping tildes for brevity.
\begin{align}
N&=F^{1/2} \\
b^1&= -\frac12\ln \frac{\lambda}{F}\\
b^2&= -\frac12 \ln \dfrac{F r^2 t^{2p_1+2p_3}}{\lambda} \\
b^3&= -\frac12\ln (Ft^{2p_1})\\
n_1&= \frac{Fkt^{2p_3}}{\lambda} - \beta_2 \\
n_2&=0\\
n_3&=0.
\end{align}
Its $\psi$-$Z$ area element
\be
\mathcal{A}=rt^{p_1+p_3}
\ee
is the same as the seed solution's, and is always expanding. Its volume element
\be
\mathcal{V}=rt \sqrt{F}
\ee
is different from the seed solution's and is not always expanding. This means the Hubble scalar $H$ can become negative for some parameter values, and Hubble-normalised variables would blow up. 
In this case we use $\beta$-normalisation, which is based on the ever-expanding area element~\cite{art:vEUW2002}.

\section{The initial goal -- late-time permanent spikes}

\begin{figure}[t]
	\begin{center}
	\includegraphics[width=8cm]{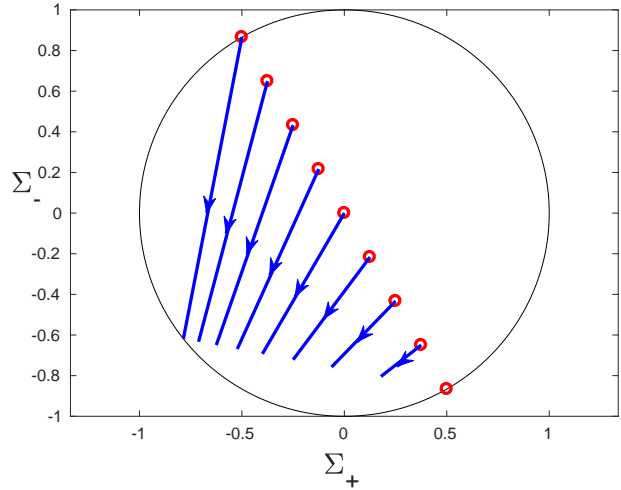}
	\end{center}
	\caption{State space orbits projected on the $(\Sp,\Sm)$ plane for various values of $\Spo$ for $k=0$, $\omega_0\neq0$. 
	A circle represents the orbit along $r=0$, which is a fixed point. $r\neq0$ orbits move away from these fixed points as $t$ increases, mimicking the orbits of Taub solutions.}	
	\label{k0orbits}
\end{figure}

For the special case $k=0$ (that is, the KVF is purely rotational), $(\Sp,\Sm)$ reduces to
\be
\label{SpsmkzeroG}
	\Sp=-\frac{\Spo+f}{2+f},\quad \Sm=\frac{\sqrt{3}(\Spo-f)}{2+f},
\ee
where 
\be
\label{fkzeroG}
	f=t(\ln F)_t=\frac{4p_1r^4t^{4p_1}}{r^4t^{4p_1}+\omega_{0}^2},\quad p_1=\frac13(1+\Spo).
\ee
Along $r=0$, we have $f=0$ and
\be
	(\Sp,\Sm)=\left(-\frac12\Spo,\frac{\sqrt{3}}{2}\Spo\right).
\ee
Along $r\neq0$, provided that $p_1\neq0$ $\Leftrightarrow$ $\Spo>-1$, we have
\be
	f\rightarrow 
		\begin{cases}
		0  &\text{as $t\rightarrow 0$}\\
		4p_1  &\text{as $t\rightarrow \infty$},
		\end{cases}
\ee
so there is a late-time permanent spike along the cylindrical axis $r=0$, for the case $k=0$, $-1<\Spo \leq 1$, $\omega_0 \neq 0$.
Figure~\ref{k0orbits} shows the state space orbits projected on the $(\Sp,\Sm)$ plane for various values of $\Spo$ for $k=0$, $\omega_{0}\neq0$. 
The $r=0$ orbits are fixed points, while the $r\neq0$ orbits move away from these fixed points as $t$ increases, 
mimicking the orbits of Taub (Bianchi type II) solutions~\cite[page 136]{book:WainwrightEllis1997}.

Following $\cite{art:Lim2008}$, we obtain the coordinate and physical radii of the spike:
\begin{align}
	\text{coordinate radius}&=|\omega_0|^\frac12t^{-p_1},\\
	\text{physical radius} &=|\omega_0|^{\frac32}\int_{0}^{1}\sqrt{1+u^4}\, \d u.
\end{align}
i.e. the physical radius of the spike is time-independent. 

This is the first generated solution with a late-time permanent spike, and the first generated solution with a spike along a line.%
\footnote{Such features can also be achieved through silent LTB and Szekeres models~\cite{art:ColeyLim2014} without using solution-generating transformations.}
The spike produces an overdensity along the axis at late times, which is conducive to large scale structure formation. 
Thus the generated solution can serve as a prototypical model for formation of galactic filaments along web-like strings.

What is special about the rotational KVF? Its length vanishes along the rotation axis. 
As the universe expands, the length squared of the KVF $\lambda$ typically increases and dominates the magnitude of its vorticity $\omega$, 
as seen in~\cite{art:Lim2015,art:ColeyLim2016,art:Coleyetal2016,art:Gregorisetal2017}.
For a rotational KVF, the rotation axis is the exceptional points where this does not happen. 
This creates a discontinuous limit for $f$ at late times -- a late-time permanent spike.

\section{A new technique to analyse the transient dynamics}

While the $k=0$ case (rotational KVF) has simple dynamics, the $k\neq0$ case (mixed KVF) has rich transient dynamics which requires the introduction of a new technique to analyse them.
We have
\be
\label{GSigma}
	\Sp=-\frac{\Spo+f}{2+f},\quad \Sm=\frac{\sqrt{3}(l-f-\frac13(2-\Spo))}{2+f},
\ee
where $f=t(\ln F)_t$, $l=t(\ln \lambda)_t$.
If $f$ becomes less than $-2$, then it is more appropriate to use $\beta$-normalised\footnote{Analogous to Hubble-normalisation, $\beta$-normalisation is based on the area element $\mathcal{A}$.
$\beta = \frac12 N^{-1} \partial_t \ln \mathcal{A}$, and is related to $H$ through $\beta = H + \sigma_+$~\cite{art:vEUW2002}.}
 $(\Sp,\Sm)$, which are
\be
	\Sp=-\frac{\Spo+f}{2-\Spo},\quad \Sm=\frac{\sqrt{3}(l-f)}{2-\Spo}-\frac{1}{\sqrt{3}}.
\ee
$l$ is a coordinate effect, so we focus on $f$.
$f$ consists of terms involving $\lambda$ and terms involving $\omega$. 
$\lambda$ contains two different power terms, $t^{2p_1}$ and $t^{2p_3}$, while $\omega$ contains $t^{1+p_3}$ and a time-independent term. 
We group them into four terms on the basis of the power of $t$:
\begin{equation}
\label{transient_division}
	T_1= r^2t^{2p_1}, \quad T_2=k^2t^{2p_3}, \quad T_3= \frac{2k}{1+p_3}t^{1+p_3}, \quad T_4= k\Spo r^2+\omega_{0}.
\end{equation}

\begin{figure}[t]
        \begin{center}
                \includegraphics[width=6cm]{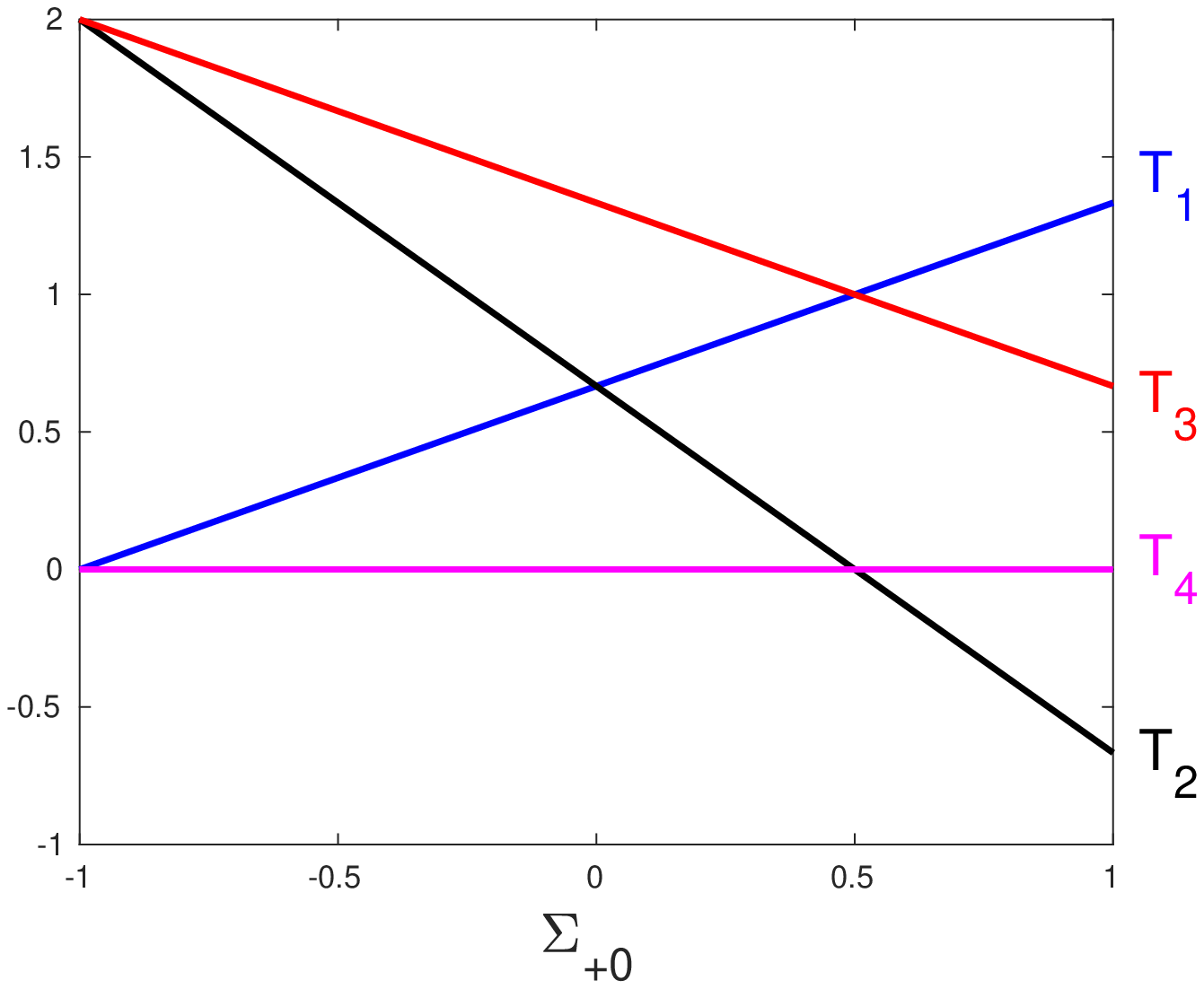}
        \end{center}
        \caption{Power of $t$ of the terms in (\ref{transient_division}) against $\Spo$.}
        \label{Texponent}
\end{figure}

Figure~\ref{Texponent} plots the power of $t$ of each term in (\ref{transient_division}) against the parameter $\Spo$. In general, the four powers are distinct, except for 3 special values of 
$\Spo$. For $\Spo=-1$, there are two distinct powers; for $\Spo=0$, three distinct powers; and for $\Spo=\frac12$, two distinct powers. The term with the largest power of $t$ dominates at late
times; the term with smallest power of $t$ dominates at early times; and the terms with intermediate power of $t$ may or may not dominate for a finite time interval, depending on how big their
coefficient is.

Expressed in terms of $T_1$, $T_2$, $T_3$, $T_4$,
\be
\label{fGeneralT}
	f=\frac{2(T_1+T_2)(2p_1T_1+2p_3T_2)+2(T_3+T_4)(1+p_3)T_3}{(T_1+T_2)^2+(T_3+T_4)^2}.
\ee
Observe that
\be
	f \approx 
		\begin{cases}
		4p_1	&\text{when $T_1$ dominates}\\
		4p_3	&\text{when $T_2$ dominates}\\
		2(1+p_3) &\text{when $T_3$ dominates}\\
		0	&\text{when $T_4$ dominates}.
		\end{cases}
\ee
That is, $f$ is approximately twice the value of the power of the dominant term. Furthermore the powers depend only on the parameter $\Spo$. Its independence of coordinates gives the graph of $f$ 
a cascading appearance. An equilibrium state corresponds to a dominant term. Therefore, there are up to 4 distinct equilibrium states for general $\Spo$; 3 for $\Spo=0$ and 2 for $\Spo=-1$ and
$\Spo=\frac12$. The value of $f$ at successive equilibrium states is strictly increasing in time. Among the four values, $4p_3$ is negative for $\frac12<\Spo\leq 1$, with a minimum value of $-\frac{4}{3}$
at $\Spo=1$, which is still greater than $-2$, so the Hubble scalar $H$ is positive at each equilibrium state. But we will see later that $f$ can become less than $-2$ during so-called overshoot
transitions.

We define the transition time between two equilibrium states or dominant terms to be the time when both terms are equal in magnitude. 
For example, solving $T_1^2=T_2^2$ for $t$ yields the transition time
\begin{equation}
	t_{12}=\left(\frac{k^2}{r^2}\right)^\frac{1}{2\Spo}.
\end{equation}
Comparing the transition times will determine how many transitions an observer with fixed $r$ undergoes.
The coefficients of $T_1$ and $T_4$ have spatial dependence. They can even vanish for certain worldline ($r=0$ for $T_1$, and $r=\sqrt{\frac{-\omega_0}{k\Spo}}$ for $T_4$, provided that 
$\frac{\omega_{0}}{k \Spo}\leq0$), which create spikes along these worldlines. The spikes are called transient if the term dominates for an intermediate, finite time interval. Some transition
times have spatially dependence as a consequence of the spatially dependent coefficient. This means there are inhomogeneities in transition times except $t_{23}$.

The transition time between two dominant terms can be regarded as roughly the boundary between the two corresponding equilibrium states. We say ``roughly" because the transition is a smooth, 
continuous process, so there is no sharp boundary. If a transition time has spatial dependence, it also gives the spatial location of the boundary at a fixed time. The spacetime is partitioned
into regions of equlibrium states, separated by transition times. When viewed at a fixed time, we can regard space as being partitioned into cells of equilibrium states, separated by walls (around
which spatial gradient is large). If two walls are near each other, we see a narrow cell.
The neighbourhood of the narrow cell shall be called a spike if certain additional conditions are met. We will discuss these conditions in Section~\ref{sec:example3}.

We now give a number of examples to show the various features. 

\subsection{Example 1}

For the case $-1<\Spo<0$, Figure \ref{Texponent} gives the ordering $T_4$,  $T_1$, $T_2$, $T_3$, in increasing power of $t$. 
We have up to 4 distinct equilibrium states, and along general worldlines there are 4 possible sequences of dominant equilibrium states, which we shall refer to as scenarios:
\begin{enumerate}
\item  $T_{4}$ $\longrightarrow$ $T_{1}$ $\longrightarrow$ $T_{2}$ $\longrightarrow$ $T_{3}$
\item  $T_{4}$ $\longrightarrow$ $T_{2}$ $\longrightarrow$ $T_{3}$
\item  $T_{4}$ $\longrightarrow$ $T_{1}$ $\longrightarrow$ $T_{3}$
\item  $T_{4}$ $\longrightarrow$ $T_{3}$.
\end{enumerate}
There are two special worldlines where a term vanishes. The first one is $r=0$, where $T_1$ vanishes. The possible scenarios along this worldline are:
\begin{enumerate}
\item  $T_{4}$ $\longrightarrow$ $T_{2}$ $\longrightarrow$ $T_{3}$
\item  $T_{4}$ $\longrightarrow$ $T_{3}$,
\end{enumerate}
which are qualitatively the same as scenarios 2 and 4 above. Because of this, $r=0$ is not really special. 
This suggests that transient spikes do not occur along a special worldline, but rather require a scenario in which some intermediate term is always sub-dominant.
The second special worldline is $r=\sqrt{\frac{-\omega_{0}}{k \Spo}}$, where $T_4$ vanishes, giving an early-time permanent spike.
The possible scenarios along this worldline are:
\begin{enumerate}
\item  $T_{1}$ $\longrightarrow$ $T_{2}$ $\longrightarrow$ $T_{3}$
\item  $T_{1}$ $\longrightarrow$ $T_{3}$.
\end{enumerate}
The two special worldlines coincide if $\omega_{0}=0$. In this case the only possible scenarios along this worldline is
\[
	T_{2} \longrightarrow T_{3}.
\]

\begin{figure}[t]
        \begin{center}
                \includegraphics[width=6cm]{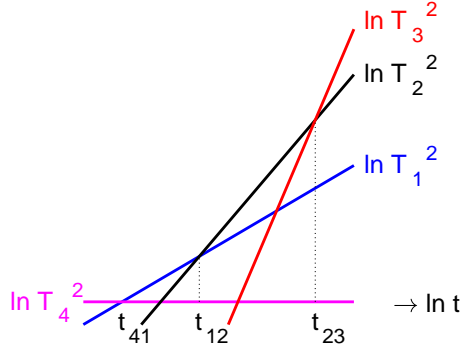}
        \end{center}
        \caption{Qualitative plot of the log of each term squared against $\ln t$ for the scenario $T_{4} \longrightarrow  T_{1} \longrightarrow   T_{2}  \longrightarrow  T_{3}$.}
        \label{d4123}
\end{figure}

\begin{figure}[t]
	\begin{center}
	\begin{subfigure}[t]{6cm}
                \includegraphics[width=6cm]{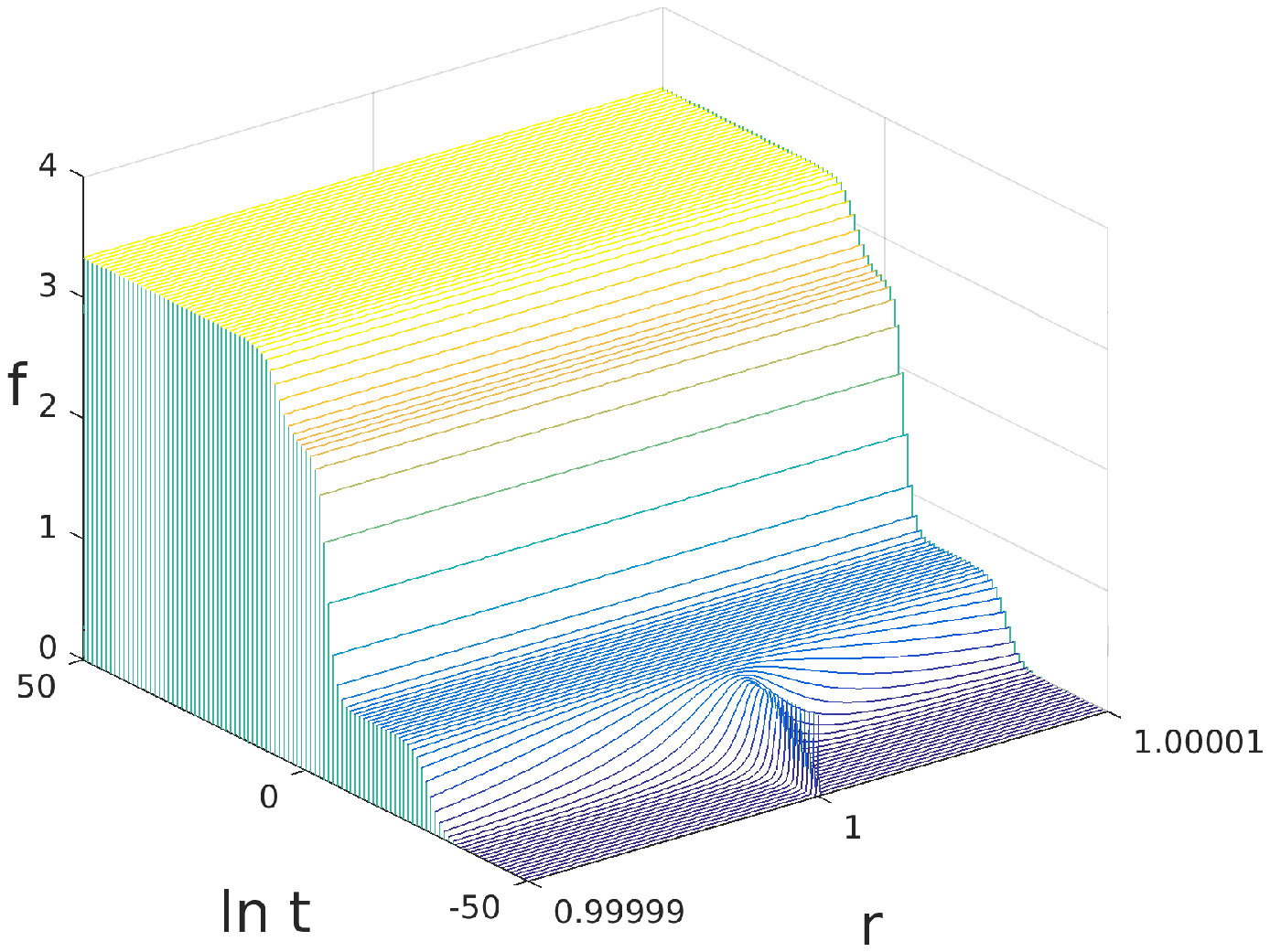}
	\end{subfigure}
	\begin{subfigure}[t]{6cm}
		\includegraphics[width=6cm]{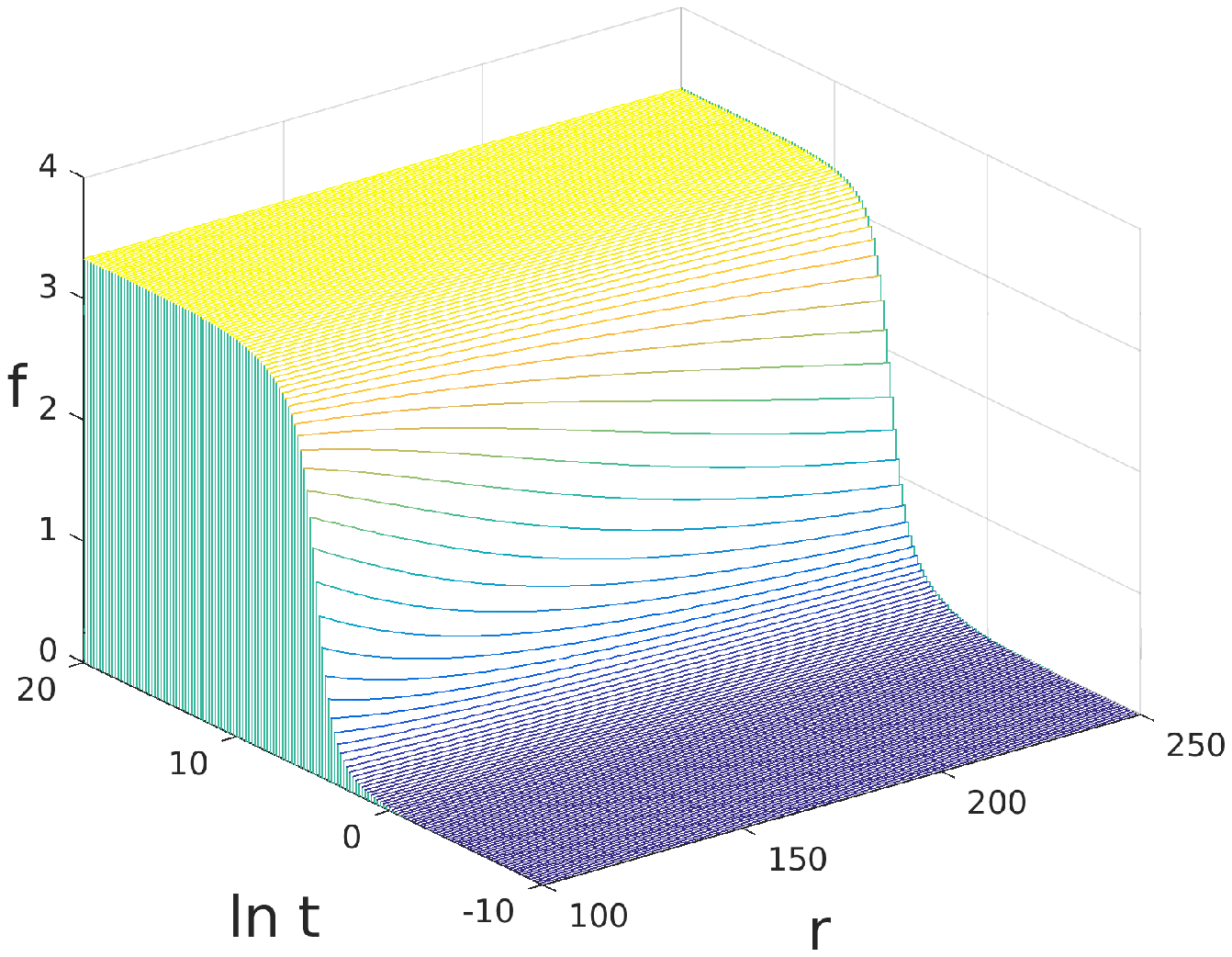}
	\end{subfigure}
	\end{center}
	\caption{$f$ against $\ln t$ and $r$, showing the intervals where the scenario 
		$T_{4} \longrightarrow  T_{1} \longrightarrow   T_{2}  \longrightarrow  T_{3}$ occurs. The interval $0.99999<r<1.00001$ shows four distinct states along $r\neq1$.
		The interval $100<r<250$ shows only two visible distinct states because the transition times are too close together.}
	\label{f4123}
\end{figure}

We now introduce a useful diagram. From (\ref{transient_division}), we see that the logarithm of the square of each term is a linear function of $\ln t$. 
Figure \ref{d4123} shows a qualitative plot of the log of each term squared against $\ln t$ for the scenario
\[
	T_{4} \longrightarrow  T_{1} \longrightarrow   T_{2}  \longrightarrow  T_{3}.
\]
The plot is useful for determining the order of the transition times.
It is clear from the diagram that the transition times
\be
	t_{41}=\left(\frac{|k \Spo r^2 +\omega_{0}|}{r^2}\right)^{\frac{1}{2p_1}},\quad t_{12}=\left(\frac{k^2 }{r^2}\right)^{\frac{1}{2\Spo}}, \quad
	t_{23}=\left(\frac{|k|(2- \Spo)}{3}\right)^{\frac{1}{2p_1}}
\ee
must satisfy the condition
\be
\label{4stageconditionm05}
	t_{41}<t_{12}<t_{23}
\ee
in this scenario.
The condition then determines the $r$ intervals (the worldlines) where the scenario occurs.
$t_{41}<t_{12}$ implies
\be
\label{eq1hash}
	|k \Spo r^2 +\omega_{0}| < \left(\frac{|k|^{p_1}}{r^{p_3}}\right)^\frac{2}{\Spo},
\ee
which gives one or more intervals of $r$.
$t_{12}<t_{23}$ gives an upper bound on $r$:
\begin{equation}
\label{eq1hashr}
	r<|k|\left(\frac{3}{|k|(2-\Spo)}\right)^{\frac{\Spo}{2 p_1}}.
\end{equation}
So the condition (\ref{4stageconditionm05}) restricts $r$ to one or more intervals.
As a concrete example, take the parameter values
\be
	\Spo=-0.5,\quad k=10,\quad \omega_{0}=5.
\ee
(\ref{eq1hash}) can be solved numerically to give the intervals
\be
\label{eq2hash}
	0.9794<r<1.0226\ \text{and} \ 111.7900<r.
\ee
Note that $r=1$ is the second special worldline, so it must be excluded from this scenario.
(\ref{eq1hashr}) gives $r<240.5626$. Together, the scenario occurs for the intervals
\be
	0.9794<r<1,\quad  1<r<1.0226 \ \text{and} \ 111.7900<r<240.5626.
\ee
We plot $f$ against $\ln t$ and $r$ in Figure~\ref{f4123}, showing the intervals where the scenario occurs. The interval $0.99999<r<1.00001$ shows four distinct states along $r\neq1$.
Along the cylindrical shell $r=1$, there is a permanent spike at early times.
The interval $100<r<250$ shows only two visible distinct states because the transition times are too close together. 
So if transition times are too close together, we see fewer visible distinct state than the actual number of states predicted by the scenario.

\begin{figure}[t]
        \begin{center}
                \includegraphics[width=6cm]{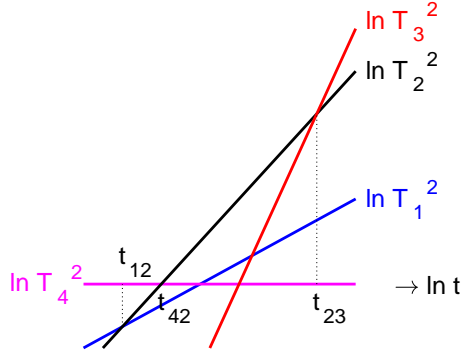}
        \end{center}
        \caption{Qualitative plot of the log of each term squared against $\ln t$ for the scenario $T_{4} \longrightarrow   T_{2}  \longrightarrow  T_{3}$.}
        \label{d423}
\end{figure}

\begin{figure}[t]
        \begin{subfigure}[t]{6cm}
                \includegraphics[width=6cm]{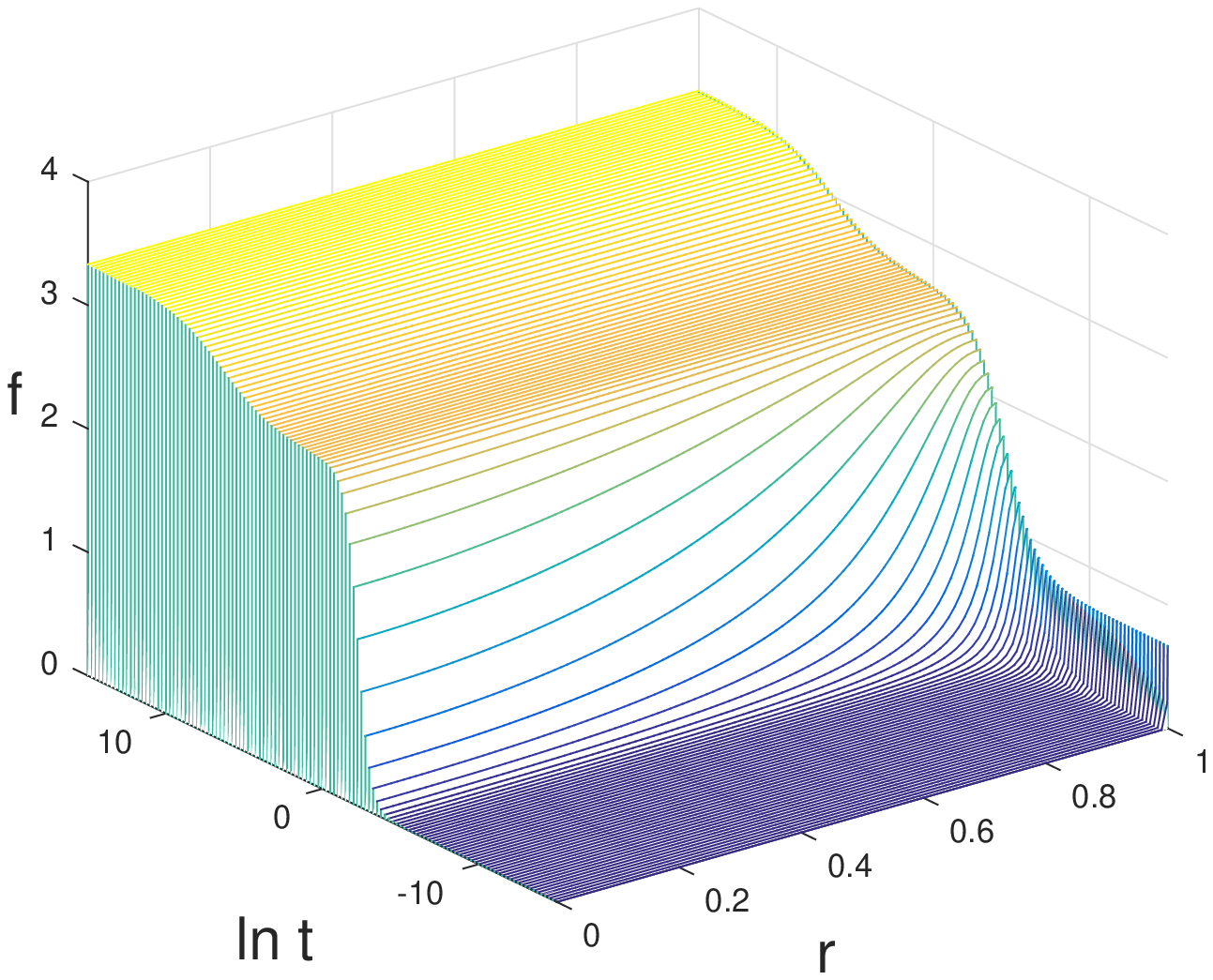}
        \end{subfigure}
        \begin{subfigure}[t]{6cm}
                \includegraphics[width=6cm]{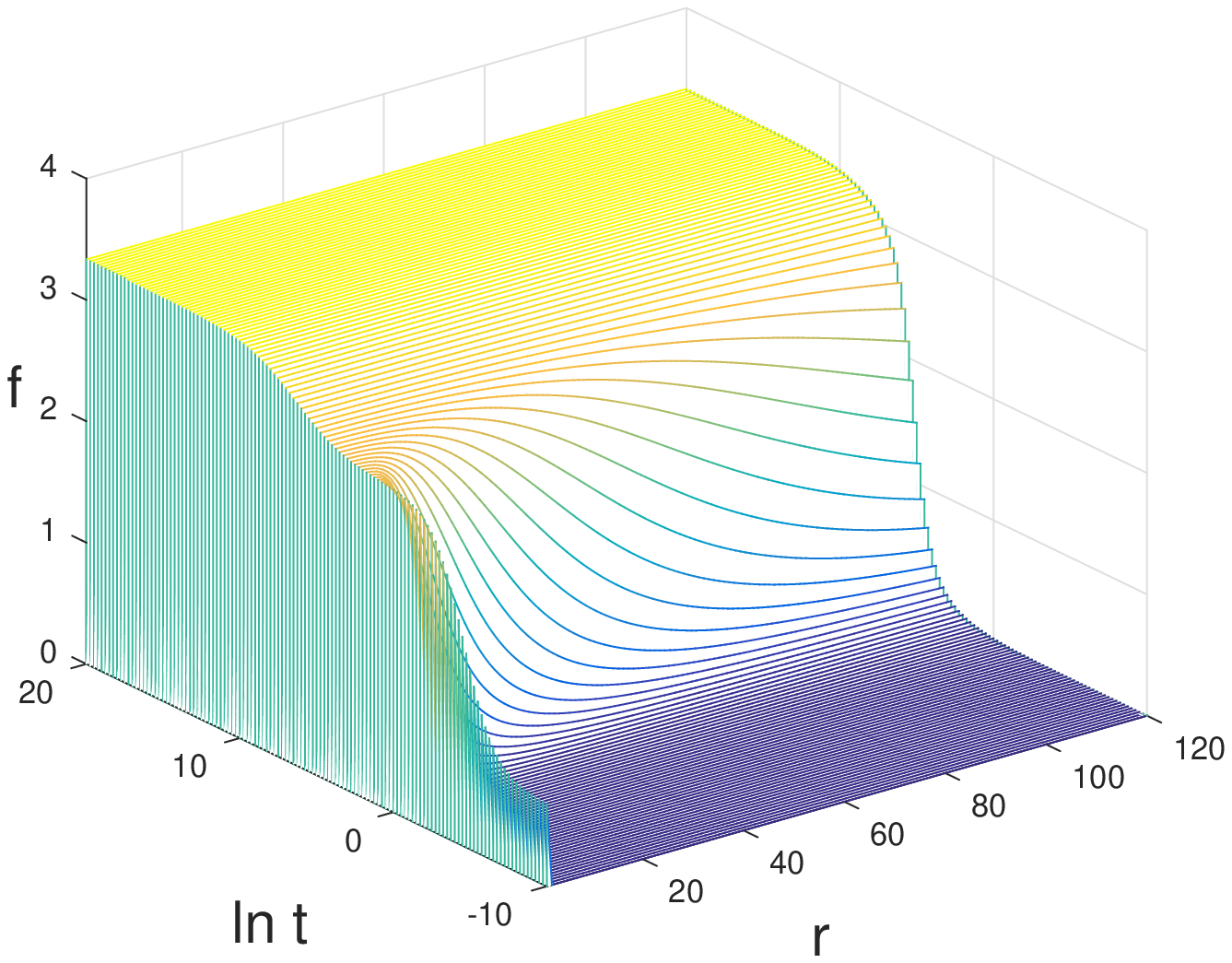}
        \end{subfigure}
	\caption{$f$ against $\ln t$ and $r$, showing the intervals where the scenario
                $T_{4} \longrightarrow   T_{2}  \longrightarrow  T_{3}$ occurs. The interval $0\leq r<0.9794$ shows three distinct states. 
		The interval $1< r<120$ shows three 3 distinct states for small $r$ which fade away to two visible states as the transition times become closer together as $r$ increases.}
	\label{f423}
\end{figure}

What happens in other intervals of $r$?
From (\ref{eq2hash}), we know that $t_{41}$ becomes greater than $t_{12}$ for values of $r$ just beyond the boundaries. 
From the diagram in Figure \ref{d4123}, this happens if the graph of $\ln T_1^2$ becomes too low, as shown in Figure \ref{d423}.
Now, the diagram in Figure \ref{d423} shows the scenario
\be
	T_{4} \longrightarrow T_{2} \longrightarrow T_{3},
\ee
with transition times
\be
	t_{42}=\left(\frac{|k \Spo r^2 +\omega_{0}|}{k^2}\right)^{\frac{1}{2p_3}},\quad t_{23}=\left(\frac{|k|(2- \Spo)}{3}\right)^{\frac{1}{2p_1}}.
\ee
They must satisfy the condition
\be
\label{3stage1conditionm05}
	t_{12}<t_{42}<t_{23}.
\ee
The condition $t_{12}<t_{42}$ is equivalent to $t_{12}<t_{41}$, so it gives (\ref{eq1hash}) with the opposite inequality direction:
\be
\label{eq3hash}
	|k \Spo r^2 +\omega_{0}| > \left(\frac{|k|^{p_1}}{r^{p_3}}\right)^\frac{2}{\Spo}.
\ee
$t_{42}<t_{23}$ implies
\begin{equation}
\label{eq4hash}
	|k \Spo r^2 +\omega_{0}|<|k|^{\frac1{p_1}}\left(\frac{2-\Spo}{3}\right)^\frac{p_3}{p_1},
\end{equation}
which gives rise to one interval of $r$. Together, the condition restricts $r$ to one or more intervals. (\ref{eq3hash}) gives the intervals
\be
	r<0.9794 \ \text{and}\ 1.0226<r<111.7900,
\ee
while
(\ref{eq4hash}) gives the interval $r<310.5666$. Together, the scenario occurs for the intervals
\be
\label{423intervals}
	0\leq r<0.9794 \ \text{and}\ 1.0226<r<111.7900.
\ee
We plot $f$ against $\ln t$ and $r$  on these intervals showing the scenario $T_{4} \longrightarrow   T_{2}  \longrightarrow  T_{3}$ in Figure~\ref{f423}.
The interval $0\leq r<0.9794$ shows three distinct states.
The interval $1< r<120$ shows three 3 distinct states for small $r$ which fade away to two visible states as the transition times become closer together as $r$ increases.

\begin{figure}[t!]
        \begin{center}
                \includegraphics[width=6cm]{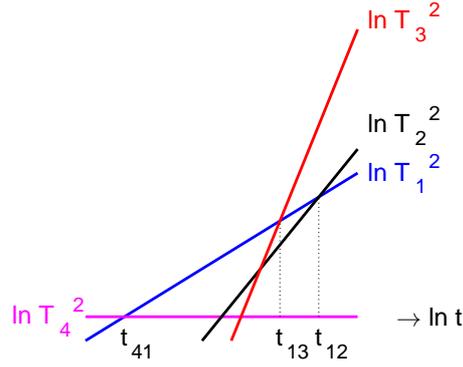}
        \end{center}
        \caption{Qualitative plot of the log of each term squared against $\ln t$ for the scenario $T_{4} \longrightarrow   T_{1}  \longrightarrow  T_{3}$.}
        \label{d413}
\end{figure}

\begin{figure}[t]
        \begin{center}
                \includegraphics[width=6cm]{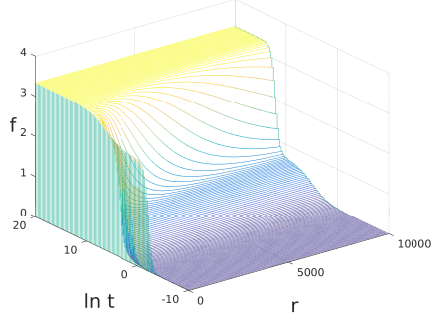}
        \end{center}
        \caption{$f$ against $\ln t$ and $r$, 
                showing the scenario $T_{4} \longrightarrow T_{1} \longrightarrow T_{3}$ for $r>240.5626$.}
        \label{f413}
\end{figure}

To complete the example, we now look at what happens beyond $r=240.5626$, where $t_{12}$ becomes larger than $t_{23}$. 
From the diagram in Figure \ref{d4123}, this happens if the graph of $\ln T_2^2$ becomes too low, as shown in Figure \ref{d413}. 
Now, the diagram in Figure \ref{d413} shows the scenario
\be
	T_{4} \longrightarrow T_{1} \longrightarrow T_{3},
\ee
with transition times
\be
	t_{41}=\left(\frac{|k \Spo r^2 +\omega_{0}|}{r^2}\right)^{\frac{1}{2p_1}},\quad t_{13}=\left(\frac{r^2(2- \Spo)}{3|k|}\right)^{\frac{1}{2p_3}}.
\ee
They must satisfy the conditions
\be
\label{3stage2conditionm05}
	t_{41}<t_{13}<t_{12}.
\ee
The condition $t_{13}<t_{12}$ is equivalent to $t_{23}<t_{12}$, so it gives (\ref{eq1hashr}) with the opposite inequality direction, a simple lower bound
\be
\label{eq7.15}
	r>|k|\left(\frac{3}{|k|(2-\Spo)}\right)^{\frac{\Spo}{2 p_1}}.
\ee
The condition $t_{41}<t_{13}$ implies
\be
\label{eqstar}
	|k \Spo r^2 +\omega_{0}|<r^\frac{1+p_3}{p_3}\left(\frac{2-\Spo}{3|k|}\right)^\frac{p_1}{p_3},
\ee
which restricts $r$ to one or more intervals.
(\ref{eq7.15}) gives the interval $r>240.5626$, while (\ref{eqstar}) gives the intervals
\be
	0.9514<r<1.0605 \ \text{and}\ r>86.5794.
\ee
Together, the scenario occurs for $r>240.5626$.
We plot $f$ against $\ln t$ and $r$ showing 3 distinct states with a lower bound on $r$ in Figure \ref{f413}.

\begin{figure}[t]
	\begin{center}
	\begin{subfigure}[t]{6cm}
		\includegraphics[width=6cm]{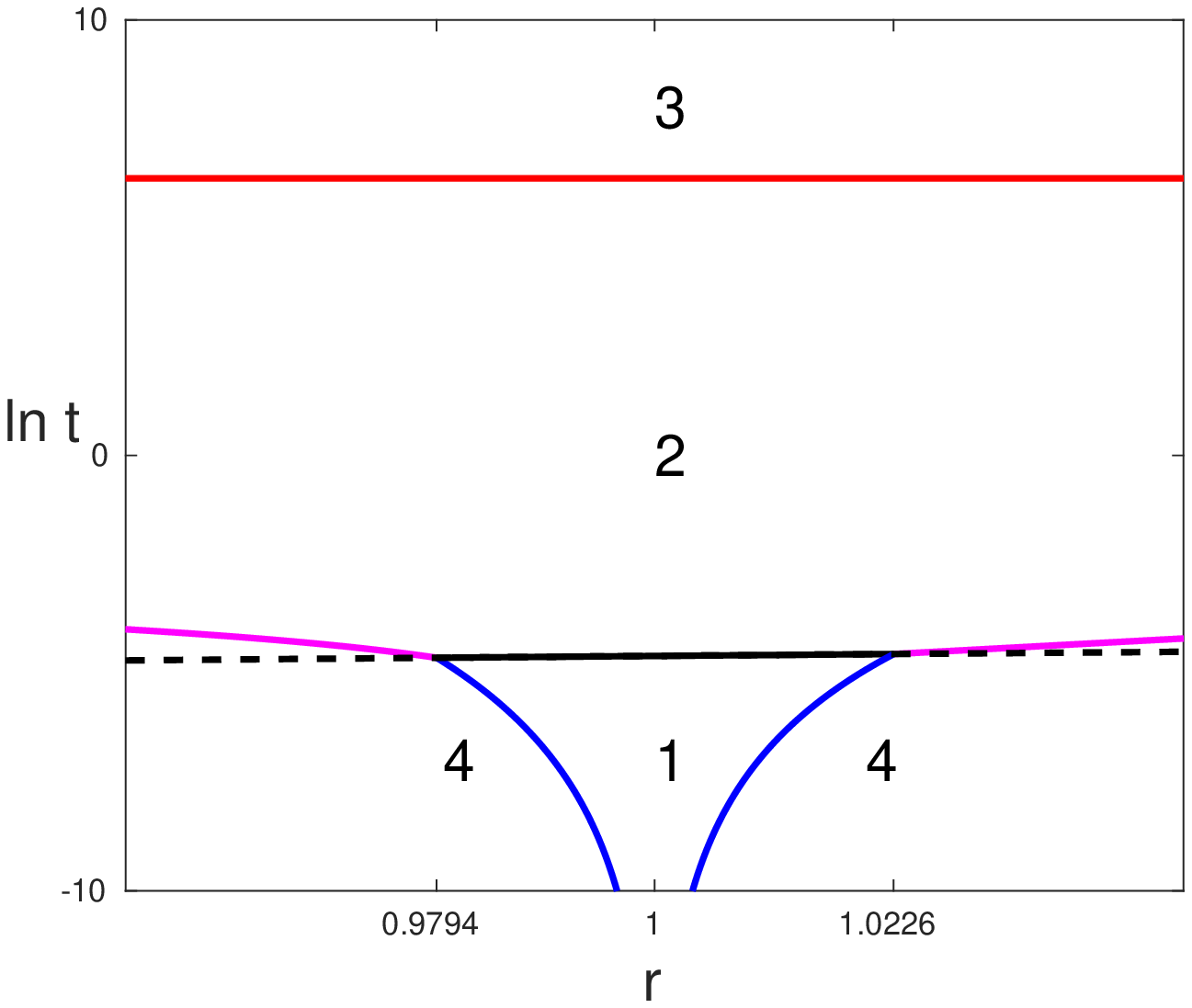}
	\end{subfigure}
	\begin{subfigure}[t]{6cm}
                \includegraphics[width=6cm]{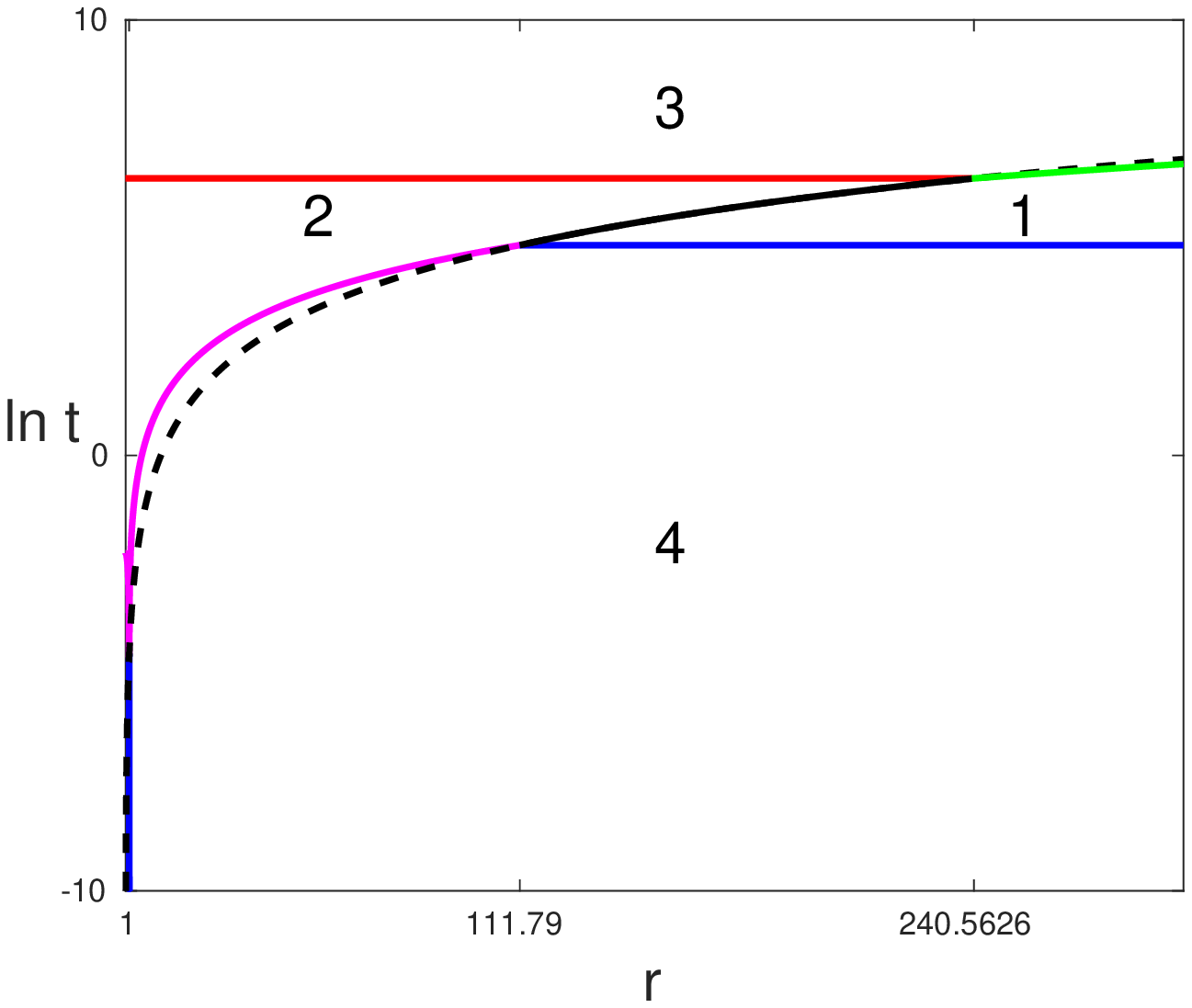}
        \end{subfigure}
	\end{center}
        \caption{Plot of the cells and transition times, showing the different scenarios along each fixed $r$.
		Each cell is labelled with the index of the dominant term. Dashed line indicates the transition time $t_{12}$ for $l$.}
        \label{cell1}
\end{figure}

\begin{figure}[t]
        \begin{center}
                \includegraphics[width=12cm]{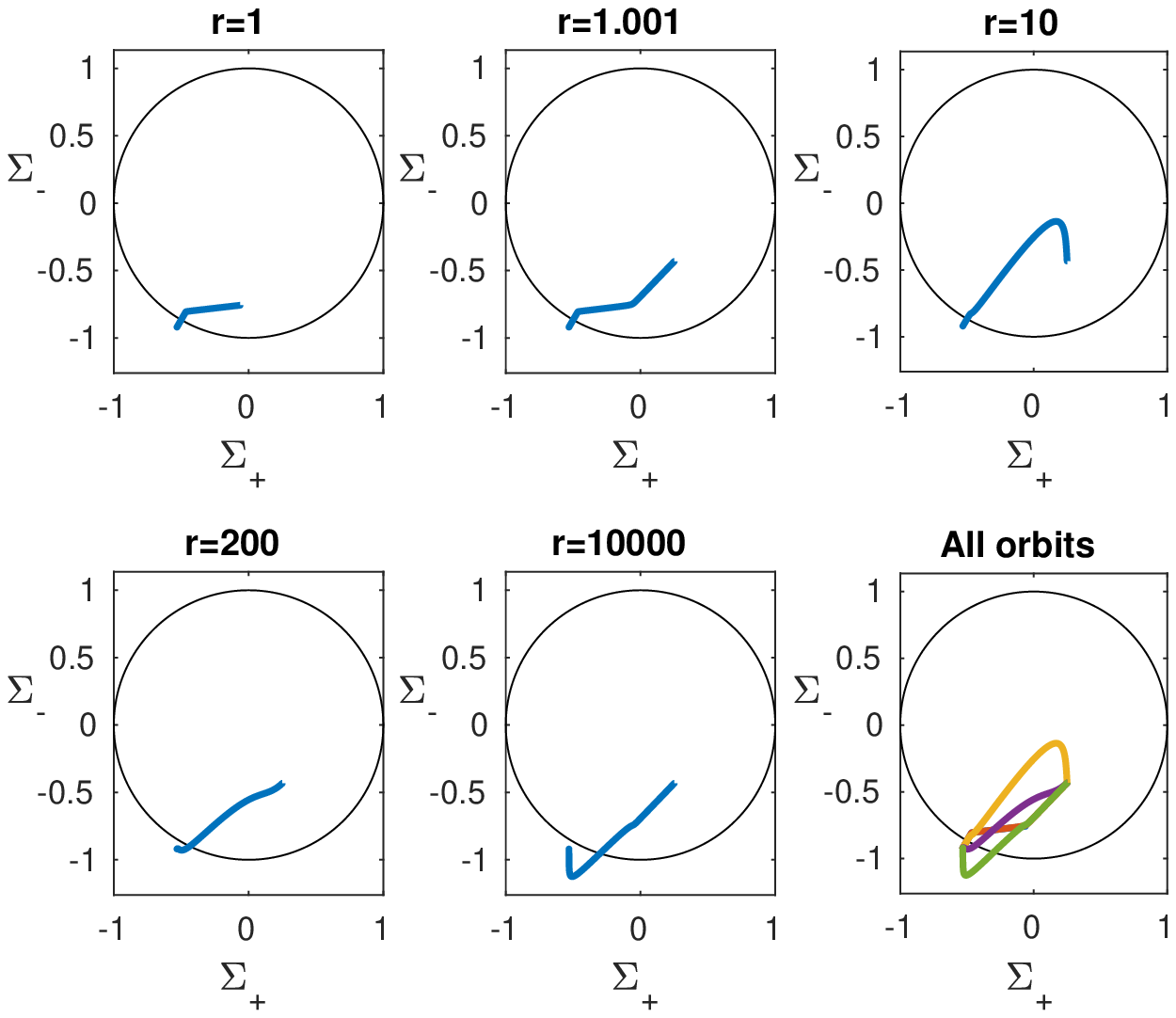}
        \end{center}
        \caption{State space orbits along representative worldlines.}
        \label{orbits1}
\end{figure}

This completes the scenarios in this example. We summarise them in another useful diagram, 
where we plot the transition times of each scenario, and label the dominant term in each cell. See Figure \ref{cell1}.
Are there transient spikes? Not at first sight.
Equation~(\ref{423intervals}) gives the two $r$ intervals where the worldlines undergo the scenario $T_4 \longrightarrow T_2 \longrightarrow T_3$.
The first one could be called a transient spike, even though it looks wide in comparison to the narrow permanent spike around $r=1$.
The second one however is too wide to be considered a spike.
Perhaps we should shift our focus from spikes to cells of various length scales.
Very narrow cells are obvious candidates for spikes, but the visual distinction fades for wider cells.

Plotting the state space orbits reveals that the solution is future asymptotic to a state with the following values:
\be
	\Sp = -\frac{17}{32},\quad \Sm = -\frac{17\sqrt{3}}{32},\quad \Sigma_{12}=0,\quad \Omega = \frac{27}{256},\quad \Omega_k= -\frac{15}{64},
\ee
a presently unidentified non-vacuum state with negative spatial curvature parameter $\Omega_k$.
See Figure~\ref{orbits1}. 
Figure~\ref{cell1} shows that the transition time of $l$, which is $t_{12}$, happens to be close to a transition time of $f$ in this example, 
so the state space orbits in Figure~\ref{orbits1} do not have a distinctive vertical segments like in Figure~\ref{seedorbits}.

\section{Example 2}

\begin{figure}[t]
	\begin{center}
	\begin{subfigure}[t]{6cm}
                \includegraphics[width=6cm]{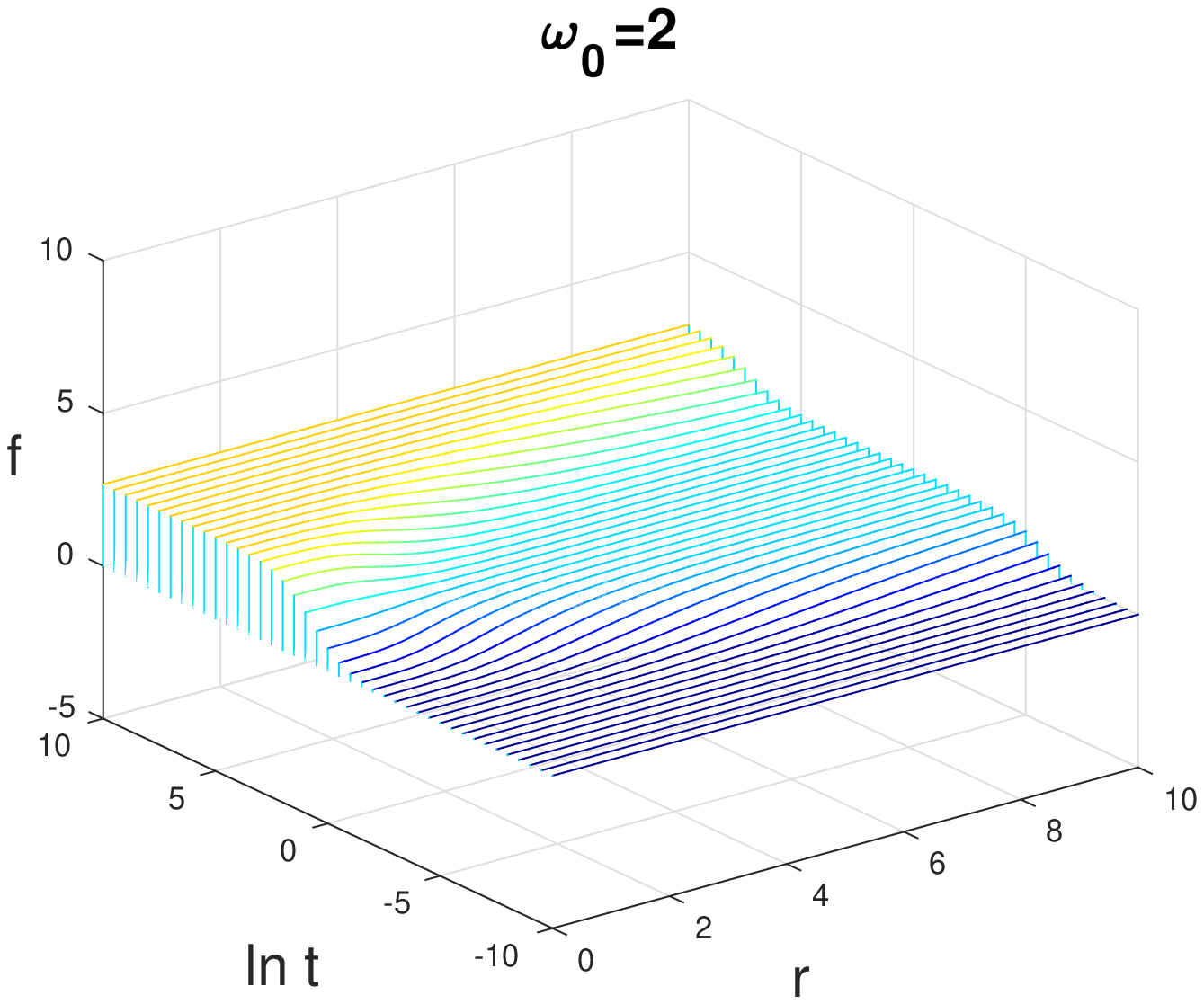}
        \end{subfigure}
        \begin{subfigure}[t]{6cm}
                \includegraphics[width=6cm]{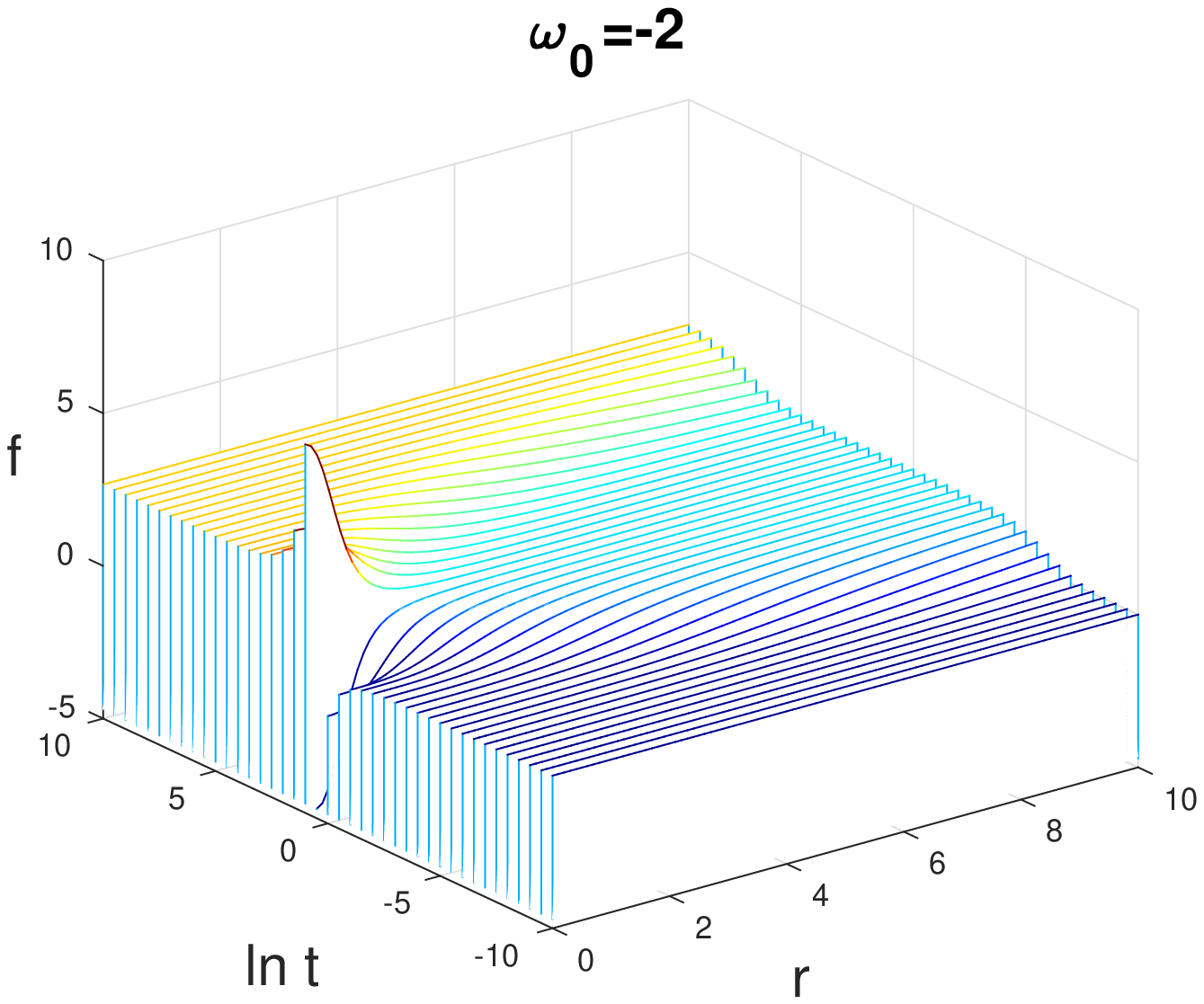}
        \end{subfigure}
	\end{center}
	\caption{$f$ against $\ln t$ and $r$ for Example 2, showing scenario (\ref{Scenariozero1}) for $r>1.0950$ and scenario (\ref{Scenariozero2}) for $r<1.0950$. 
		The $T_{4} \longrightarrow T_{3}$ transition is sigmoid for $\omega_{0}=2$, but has overshoots for $\omega_{0}=-2$.}
	\label{f_example2}
\end{figure}

The second example showcases a transient spike and an overshoot transition.
For $\Spo=0$, we have 
\[
	T_1= r^2t^{\frac23}, \ \ T_2=k^2t^{\frac23}, \ \ \ T_3= \frac{3k}{2}t^{\frac43},\ \  \ T_4= \omega_{0}.
\]
There are only three distinct powers of $t$, with $T_4$ dominating at early times, $T_3$ dominating at late times, and $T_1$ and $T_2$ possibly at intermediate times. 
The first scenario is the 3-state sequence
\be
\label{Scenariozero1}
	T_{4} \longrightarrow T_{1} \ \& \ T_{2}   \longrightarrow T_{3}
\ee
with transition times
\be
	t_{4\left(1\&2\right)}=\left(\frac{\omega_{0}^2}{r^4+k^4}\right)^{\frac{3}{4}},\quad t_{\left(1\&2\right)3}=\left(\frac{4(r^4+k^4)}{9k^2}\right)^{\frac{3}{4}},
\ee
which are required to satisfy the condition
\be
	t_{4\left(1\&2\right)}<t_{\left(1\&2\right)3}.
\ee
The condition gives a lower bound on $r$
\begin{equation}
	r>\left(\frac32|k\omega_{0}|-k^4\right)^\frac14.
\end{equation}
If the lower bound is positive, then for $r$ less than this we have the second scenario, the 2-state sequence
\be
\label{Scenariozero2}
	T_{4}  \longrightarrow T_{3}  
\ee
with transition time
\be
	t_{43}= \left|\frac{2\omega_{0}}{3k}\right|^{\frac{3}{4}}.
\ee
For example, given $k=0.5$ and $\omega_{0}=\pm2$, for $r>1.0950$ we have the scenario (\ref{Scenariozero1}) and for $r<1.0950$ we have the scenario (\ref{Scenariozero2}). 
We plot $f$ against $\ln t$ and $r$, showing both scenarios in Figure \ref{f_example2}.
The $T_{4} \longrightarrow T_{3}$ transition is sigmoid for $\omega_{0}=2$, but has overshoots for $\omega_{0}=-2$.

We noted earlier that $f$ has a cascading appearance. 
Despite this, $f$ can fluctuate wildly with overshoots.
Under what condition does this happen? If we examine $f$ from (\ref{fGeneralT}):
\be
\label{fGeneralT2}
	f=\frac{2(T_1+T_2)(2p_1T_1+2p_3T_2)+2(T_3+T_4)(1+p_3)T_3}{(T_1+T_2)^2+(T_3+T_4)^2},
\ee
we see that the magnitude of $f$ becomes large if the denominator becomes small due to cancellation. 
Among $T_1$, $T_2$, $T_3$, $T_4$, only $T_4$ can become negative, so cancellation is only possible if $T_4$ is negative. 
Cancellation happens when
\be
	\omega = T_3 + T_4 \approx 0.
\ee
Its effect is most prominent when cancellation occurs during the 
\be
\label{ost1}
	T_4 \longrightarrow T_3
\ee
transition in a scenario.
Heuristically, when $T_3 +T_4= \epsilon$, where $|\epsilon|$ is small, and suppose $T_1$ and $T_2$ are $o(\epsilon)$ at that instant, then \eqref{fGeneralT2} implies
\be
\label{ost3}
	f\approx \frac{2(1+p_3)}{\epsilon}T_3.
\ee
Then $f$ becomes negative in the first stage of the transition (when $\epsilon<0$), then positive in the second stage (when $\epsilon>0$). 
This produces two overshoots, whose amplitude can be large if $T_1$ and $T_2$ are much smaller than $T_3$ and $T_4$ when this happens. 
We therefore call such a transition an overshoot transition.

In Example 2, both $\omega_0 =\pm 2$ cases have a transient spike for the spacetime region $r \lesssim 2$, $-2 \lesssim \ln t \lesssim 3$.
The transient spike transition occuring around $\ln t \approx 0.7$ is sigmoid in the case $\omega_0 = 2$, and is an overshoot transition in the case $\omega_0 =-2$.
This example illustrates that a transient spike and an overshoot transition are two different phenomena, and an overshoot transition can occur inside a transient spike.

\section{Example 3}\label{sec:example3}

\begin{figure}[t]
        \begin{center}
                \includegraphics[width=6cm]{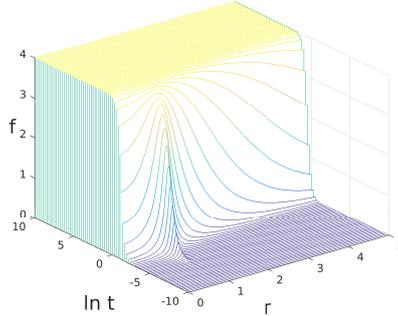}
        \end{center}
        \caption{$f$ against $\ln t$ and $r$ for Example 3. The cell is narrow near the global minimum of the transition time, but is not a transient spike.}
        \label{f_example3}
\end{figure}

The third example shows a narrow cell near the local extremum of the transition time.
Take
\be
	\Spo = -1,\quad k=10,\quad \omega_0 = 10.1.
\ee
The only scenario for the case $\Spo=-1$ is
\be
	T_{1} \ \& \  T_{4} \longrightarrow\ T_{2}\ \& \ T_{3}.
\ee
Solving the equation
\[
	T_{1}^2+T_{4}^2 = T_{2}^2+ T_{3}^2
\]
for $t$ yields the transition time
\be
t_{\left(1 \& 4\right)\left(2 \& 3\right)}= \left(\frac{(\omega_{0}-kr^2)^2+r^4}{k^2(k^2+1)}\right)^{\frac{1}{4}}.
\ee
which has a global minimum at $r=1$ for our example.
We plot $f$ against $\ln t$ and $r$ in Figure \ref{f_example3}, showing the cell becoming narrow near the global minimum of the transition time.
Does this count as a transient spike? No.
The wordlines in this example all undergo the same scenario.
In the original context where transient spikes were first named, the worldlines in a small neighbourhood undergo a different scenario than what worldlines further away undergo. 
Adding this criterion rules out the feature in this example as a transient spike.

The examples have led us to re-examine the definition for transient spikes.
Through Example 1, we realise that unlike permanent spikes, transient spikes do not occur along a special worldline, but only require a scenario where some intermediate term is always sub-dominant.
Such a scenario may occur on spatial intervals or cells with various length scales, and only the very narrow ones are visually distintive enough to be called a spike. 
Transient spikes are therefore visually less distinctive than permanent spikes.
Through Example 2, we discover a new phenomenon, an overshoot transition, which should not be confused with transient spike.
Through Example 3, we rule out certain narrow cells as transient spike, because the worldlines all undergo the same scenario.

For more examples, see~\cite{thesis:Moughal2021}. \cite{thesis:Moughal2021} shows that a late-time permanent spike forms along the cylindrical axis in the case $\frac12 < \Spo \leq 1$, $k\neq0$.
In other words, in order to generate a solution with a late-time permanent spike, it is not necessary for the KVF to be purely rotational.

\section{Conclusion}

To summarise, we have
found the first non-silent solution with a late-time permanent spike;
found the first spike along a line;
introduced a new technique to analyse a key function, $f$;
revised the description of transient spikes;
discovered and described overshoot transitions.
Late-time permanent spikes are more suitable than transient spikes and early-time permanent spikes in modelling structure formation.
Spikes along a line can be used to model formation of galactic filaments along web-like strings.
The new technique to analyse $f$ reveals the cell-like structure of inhomogeneous spacetimes, 
and the possible transition dynamics (regular sigmoid transition, overshoot transition) between cells.

We conclude by commenting on future research. Firstly, the family of exact solutions we found make up only a set of measure zero in the class of cylindrically symmetric 
solutions. How does a typical cylindirically symmetric solution evolve? To answer this question, it is necessary to conduct a numerical study of the class of cylindrically symmetric solutions, 
like the numerical study done for the class of non-OT $G_2$ vacuum solutions~\cite{art:Anderssonetal2005}. 
Secondly, we have used the rotational KVF of the LRS Jacobs solution. Exact solutions that admit a rotational KVF include the LRS Taub solution, the NUT (LRS Bianchi type VIII) solution, and the 
Taub-NUT (LRS Bianchi type IX) solution \cite[page 198]{book:WainwrightEllis1997}. It would be interesting to see what spiky solutions are generated from these solutions.
Thirdly, our exact solutions are OT $G_2$ solutions. In principle, non-OT $G_2$ solutions and $G_1$ solutions can be generated from a rotational KVF. Are there simple enough seed solutions that 
generate spiky solutions with such isometries?
Lastly, we are working on applying the new technique to the exact vacuum non-OT $G_2$ spike solution from~\cite{art:Lim2015} and the stiff spike solution from~\cite{art:ColeyLim2016}, 
to help improve the numerical simulation and matching in an upcoming paper that is an extension of~\cite{art:Limetal2009} and~\cite{art:HeinzleUgglaLim2012}.

\section*{Acknowledgements}

MZAM is supported by Pakistan's Higher Education Commission scholarship.

\bibliography{}

\begin{thebibliography}{10}

\bibitem{art:BergerMoncrief1993}
B.~K. Berger and V.~Moncrief,
\newblock Phys. Rev. D {\bf 48}, 4676 (1993).

\bibitem{art:LK63}
E.~M. Lifshitz and I.~M. Khalatnikov,
\newblock Adv. Phys. {\bf 12}, 185 (1963).

\bibitem{art:BKL1970}
V.~A. Belinskii, I.~M. Khalatnikov, and E.~M. Lifschitz,
\newblock Adv. Phys. {\bf 19}, 525 (1970).

\bibitem{art:BKL1982}
V.~A. Belinskii, I.~M. Khalatnikov, and E.~M. Lifschitz,
\newblock Adv. Phys. {\bf 31}, 639 (1982).

\bibitem{art:HeinzleUgglaLim2012}
J.~M. Heinzle, C.~Uggla, and W.~C. Lim,
\newblock Phys. Rev. D {\bf 86}, 104049 (2012), arXiv:1206.0932.

\bibitem{art:ColeyLim2012}
A.~A. Coley and W.~C. Lim,
\newblock Phys. Rev. Lett. {\bf 108}, 191101 (2012), arXiv:1205.2142.

\bibitem{art:LimColey2014}
W.~C. Lim and A.~A. Coley,
\newblock Class. Quant. Grav. {\bf 31}, 015020 (2014), arXiv:1311.1857.

\bibitem{art:ColeyLim2014}
A.~A. Coley and W.~C. Lim,
\newblock Class. Quant. Grav. {\bf 31}, 115012 (2014), arXiv:1405.5252.

\bibitem{art:Lim2015}
W.~C. Lim,
\newblock Class. Quant. Grav. {\bf 32}, 162001 (2015), arXiv:1507.02754.

\bibitem{art:ColeyLim2016}
A.~A. Coley and W.~C. Lim,
\newblock Class. Quant. Grav. {\bf 33}, 015009 (2016), arXiv:1511.07095.

\bibitem{art:Coleyetal2016}
A.~A. Coley, D.~Gregoris, and W.~C. Lim,
\newblock Class. Quant. Grav. {\bf 33}, 215010 (2016), arXiv:1606.07177.

\bibitem{art:Gregorisetal2017}
D.~Gregoris, W.~C. Lim, and A.~A. Coley,
\newblock Class. Quant. Grav. {\bf 34}, 235013 (2017), arXiv:1705.02747.

\bibitem{art:Geroch1971}
R.~Geroch,
\newblock J. Math. Phys. {\bf 12}, 918 (1971).

\bibitem{art:Geroch1972}
R.~Geroch,
\newblock J. Math. Phys. {\bf 13}, 394 (1972).

\bibitem{art:Stephani1988}
H.~Stephani,
\newblock J. Math. Phys. {\bf 29}, 1650 (1988).

\bibitem{thesis:Moughal2021}
M.~Z.~A. Moughal,
\newblock {\em Generating spiky solutions of {E}instein field equations with
  the {S}tephani transformation},
\newblock PhD thesis, University of Waikato, New Zealand, 2021,
  arXiv:2102.09776.

\bibitem{art:HeinzleUgglaRohr2009}
J.~M. Heinzle, C.~Uggla, and N.~R\"ohr,
\newblock Adv. Theor. Math. Phys. {\bf 13}, 293 (2009), arXiv:gr-qc/0702141.

\bibitem{book:Exactsol2002}
D.~Kramer, H.~Stephani, E.~Herlt, M.~A.~H. MacCallum, and E.~Schmutzer,
\newblock {\em Exact solutions of {E}instein's field equations} ({C}ambridge
  {U}niversity {P}ress: {C}ambridge, Cambridge, 2002).

\bibitem{art:Jacobs1968}
K.~C. Jacobs,
\newblock Astrophys. J. {\bf 153}, 661 (1968).

\bibitem{book:WainwrightEllis1997}
J.~Wainwright and G.~F.~R. Ellis,
\newblock {\em Dynamical systems in cosmology} ({C}ambridge {U}niversity
  {P}ress, Cambridge, 1997).

\bibitem{art:vEUW2002}
H.~{van Elst}, C.~Uggla, and J.~Wainwright,
\newblock Class. Quant. Grav. {\bf 19}, 51 (2002), arXiv:gr-qc/0107041.

\bibitem{art:Lim2008}
W.~C. Lim,
\newblock Class. Quant. Grav. {\bf 25}, 045014 (2008), arXiv:0710.0628.

\bibitem{art:Anderssonetal2005}
L.~Andersson, H.~{van Elst}, W.~C. Lim, and C.~Uggla,
\newblock Phys. Rev. Lett. {\bf 94}, 051101 (2005), arXiv:gr-qc/0402051.

\bibitem{art:Limetal2009}
W.~C. Lim, L.~Andersson, D.~Garfinkle, and F.~Pretorius,
\newblock Phys. Rev. D {\bf 79}, 123526 (2009), arXiv:0904.1546.

\end{thebibliography}

\end{document}